\documentclass[useAMS,usenatbib]{mn2e}

\usepackage[pdftex]{graphicx}
\usepackage{amsmath}

\newcommand{\kms}{{km s$^{-1}$}}

\usepackage{relsize}

\newcommand{\hi}{%
  \relax
  \ifmmode
    \textrm{\textsc{HI}}
  \else
    \textsc{H{\smaller}I}
  \fi
}

\newcommand{\simgt}{\lower.5ex\hbox{$\; \buildrel > \over \sim \;$}}
\newcommand{\simlt}{\lower.5ex\hbox{$\; \buildrel < \over \sim \;$}}

\title[The Shape of Dark Matter Haloes II.]{The Shape of Dark Matter Haloes\\
II. The Galactus \hi Modelling \&\ Fitting Tool}
\author[S. P. C. Peters et al.]{S. P. C. Peters$^{1}$,
P. C. van der Kruit$^{1}$\thanks{For more information, please contact P.C. van der Kruit at vdkruit@astro.rug.nl.}, 
R. J. Allen$^{2}$ and K. C. Freeman$^{3}$\\
$^{1}$Kapteyn Astronomical Institute, University of Groningen, P.O.Box 800, 9700AV Groningen, the Netherlands\\
$^{2}$Space Telescope Science Institute, 3700 San Martin Drive, Baltimore, MD 21218, USA\\
$^{3}$Research School of Astronomy and Astrophysics The Australian National University, Cotter Road Weston Creek, ACT 2611,\\
Australia}

\begin{document}

\date{Accepted 2015 month xx. Received 2015 Month xx; in original form 2015 Month xx}
\pagerange{\pageref{firstpage}--\pageref{lastpage}} \pubyear{2015}

\maketitle

\label{firstpage}

\begin{abstract}
We present a new \hi modelling tool called \textsc{Galactus}. 
                The program has been designed to perform automated fits of 
disc-galaxy models to observations. 
                It includes a treatment for the self-absorption of the gas.
                The software has been released into the public domain.
                We describe the design philosophy and inner workings of the 
program.
                After this, we model the face-on galaxy NGC\,2403, using both 
self-absorption and optically thin models, 
                showing that self-absorption occurs even in face-on galaxies.
                These results are then used to model an edge-on galaxy.
                It is shown that the maximum surface brightness plateaus seen 
in Paper I of this series \citep{Peters2015A}
are indeed signs of self-absorption.
                The apparent \hi mass of an edge-on galaxy can be drastically 
lower compared to that same galaxy seen face-on.
                The Tully-Fisher relation is found to be relatively free from 
self-absorption issues.
\end{abstract}


\section{Introduction}
Modelling of the distribution and kinematics of the neutral hydrogen in 
galaxies beyond our own, and the most nearby systems M\,31 and M\,33, took 
off in the 1970s, with the addition of spectroscopic capability to the 
Westerbork Synthesis Radio Telescope \citep{Allen1974A}.
This new tool gave observers sufficient spatial and velocity resolution 
to resolve the \hi structure and kinematics for a large number of galaxies.
In edge-on galaxies, it thus became possible to trace the outer envelope 
of the position-velocity (XV) diagram in order to get the rotation curve 
of the galaxy \citep{Sancisi1979A}.
Various strategies for this exist, such as fitting for the peak intensity 
 or fitting a one-dimensional Gaussian with fixed 
velocity dispersion to the outer edge of the XV-diagram 
\citep[e.g.][]{Garcia-Ruiz2002A}.
We refer the reader to \citet{OBrien2010B} for a detailed treatment on the 
various methods for deriving the rotation curve from a XV-diagram.

For face-on galaxies, fitting the structure and kinematics was first done by 
modelling the zeroth and first-moments maps, based on the channel maps in the 
\hi data cube.
These gave rise to the famous spider diagrams, 
in which the iso-velocity contours showed the mean velocity of the gas at each 
position in the galaxy \citep{Rogstad1971A}.
With the assumption of a thin disc, it is then possible to measure the 
rotation of the gas in rings centred on the galaxy.
The position angle and inclination could be varied as function of the 
rings. Assuming a rotation curve, 
\citet{Rogstad1974A} used it to model the 
peculiar velocity field of galaxy M\,83, and found the presence of a warp. 
\citet{Bosma1978} improved this by minimizing line of sight
velocity residuals for different  rotation velocity in order to fit
a rotation curve to the data.
This method proved quite successful at modelling galaxies. It has been 
extended by many authors into many software packages 
\citep{Begeman1987,Schoenmakers1999,Spekkens2007,Krajnovic2006}. 
We refer to \citet{Jozsa2007} for a detailed discussion on tilted-ring 
fitting to velocity fields.

The ever-continuing technological improvements led to ever-higher quality 
observations, which led to ever-better resolved galaxies, both spatially as 
well as in velocity.
A natural consequence of this was the attempt to model the \hi data cube itself.
This was first attempted by \citet{Irwin1991A,Irwin1993A}, who developed the 
\textsc{Cubit} code\footnote{Which can be found at 
www.astro.queensu.ca/$\sim $irwin/}. 
\textsc{Cubit} offers the user parameterized functions, which could be 
automatically fit to the galaxy. 
The program has later been expanded to fit more properties of galaxies 
\citep{Irwin1994A}.

\begin{figure*}
\includegraphics[width=0.32\textwidth]{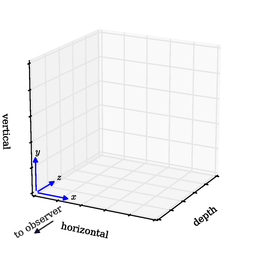}
\includegraphics[width=0.32\textwidth]{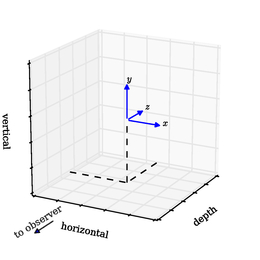}
\includegraphics[width=0.32\textwidth]{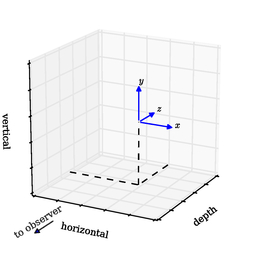}
\includegraphics[width=0.32\textwidth]{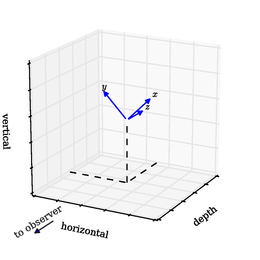}
\includegraphics[width=0.32\textwidth]{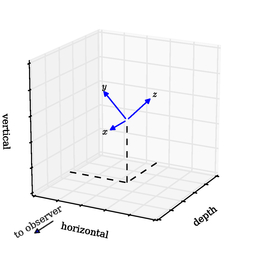}
\includegraphics[width=0.32\textwidth]{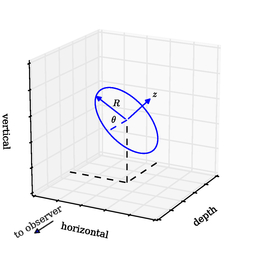}
\caption[Illustrating the coordinate system generation]{A cartoon illustrating the generation of the galaxy coordinate system. The model demonstrate the generation of an edge-on galaxy at a position angle of $135^\circ$. The boxes on each side represent the outer edges of the voxels. From top-left to bottom-right: The initial coordinate system. Centering the coordinate system (Equation \ref{eqn:cartoon1}). The shift parameters can be used to move the central position of the galaxy along the horizontal and vertical axes (Equation \ref{eqn:cartoon2}). The position angle is applied by rotating along the $z$ axis (Equation \ref{eqn:cartoon3}). The inclination is applied by rotating along the $y$ axis (Equation \ref{eqn:cartoon4}). In the final step, we switch to cylindrical coordinates. }\label{fig:cartoon}
\end{figure*}

Another software package is \textsc{GALMOD}, 
designed by T.S.~van~Albada and F.J.~Sicking, and incorporated
into the \textsc{GIPSY} data reduction system, \citep{GIPSY}.
Based on a tilted-ring geometry, \textsc{galmod} randomly projects a very 
large amount of  `clouds' in the \hi data cube.
As these clouds follow the specified kinematics and distribution, after 
sufficient samples have been drawn, the \hi data cube has been built and the 
only task remaining is scaling the cube to the desired intensity.
\textsc{Galmod} does not support automatic fitting and users thus have to 
fit a cube using the hand-and-eye strategy.

\textsc{TiRiFiC}\footnote{Currently available at 
www.astron.nl/$\sim $jozsa/tirific/index.html} is another code, 
which can automate 
the fit to a cube \citep{Jozsa2007}.
It is based on a tilted-ring model, but is expanded compared to 
\textsc{galmod}, to account for inhomogeneity in galaxy discs. 
Recently, it has used to fit features resembling spiral arms in 
edge-on galaxies \citep{Kamphuis2013A}. 

Finally, we note the \textsc{GalAPAGOS} project, which can automatically 
fit rotation curves in galaxies \citep{Wiegert2011}.
Out of all software codes currently publicly available, \textsc{TiRiFiC} 
seems to be the only one under active development. 
A fully automated three-dimensional
fitting routine  \textsc{FAT} for \textsc{Fully Automated TiRiFiC}
\citep{Kamphuis2015} and another routine, called $^{\rm 3D}$BAROLO that also 
estimates velocity dispersions \citep{Teodoro2015} has become available.

In Paper I \citep{Peters2015A}, 
we presented the \hi observations for eight edge-on galaxies.
One of the key conclusions of that paper was that self-absorption could 
well play an important role.
The effect of \hi self-absorption has long been a topic of concern, for 
example in the work by \citet{Sancisi1979A} and of \citet{Braun2009}.
The latter estimated quite significant corrections for self-absorption in the 
case of the Andromeda Nebula M\,31. 
Yet in most of the publications featuring models of neutral hydrogen in 
galaxies, the assumption remains that of an optically thin gas.

None of the above software packages supports modelling of self-absorption 
of the gas. 
This, combined with some other research questions we had concerning the 
neutral hydrogen, such as the behavior of the velocity dispersion as function 
of height above the plane,
 led us to develop a new tool for generating \hi data cubes that could treat for 
self-absorption.
Rather than using a tilted-ring model, we model the galaxy as a single plane 
on which harmonic offsets above or below the central plane represent the warps.
The code was developed in the \textsc{Python} and \textsc{C++} programming 
languages, 
 with the computationally most intensive parts implemented as multi-threaded.
The software is called \textsc{Galactus}. It has been publicly released under 
an open-source license and is listed in the Astrophysics Source Code Library 
as \citet[ascl.net/1303.018,][]{GalactusASCL}.
The code can be downloaded at sourceforge.net/p/galactus/.

The purpose of this paper is twofold. 
In Section \ref{sec:galactusdesign}, we present the inner workings of 
\textsc{Galactus}.
Section \ref{sec:notes} discusses some important features of the tool 
in more detail.
In Section \ref{sec:effectiveTspin}, we explore the effective spin 
temperature in more detail.
We test the program on face-on galaxy NGC\,2403 in Section \ref{sec:NGC2403}, 
using both an optically thin and a self-absorbing model.
In Section \ref{sec:NGC2403edgeon}, the fits from NGC\,2403 are projected to 
an edge-on galaxy to show that the maximum surface brightness profiles seen 
in Paper I should indeed have been higher.
We test how self-absorption can lower the apparent total \hi mass, depending 
on the inclination, in Section \ref{sec:effectofinclination}.
The Tully-Fisher relation
is shown to be independent of \hi self-absorption, in 
Section \ref{sec:TullyFisher}.

\section{\textsc{Galactus} Design}\label{sec:galactusdesign}
In this section, we discuss how \textsc{Galactus} generates a \hi data cube.
This generation consists of two steps.
In the first phase, the program creates a 3D model of the galaxy, while the 
second phase performs the radiative transfer.

To generate the model the software creates a three dimensional grid.
The first two axes are aligned with, and have the same lengths as, the 
horizontal and vertical  
axes of the \hi data cube channels\footnote{Note that we assume the size of the 
pixel is not frequency dependent. This is not true by default for \hi data cubes 
created using \textsc{Miriad}.}.
The third axis does not denote velocity, but represents physical depth.
The length of the third axis is the same as that of
the horizontal axis of the \hi data 
cube\footnote{When modelling only one side of a galaxy, this length is 
doubled.}.

Each position in this model is a three dimensional pixel, also known as a 
volumetric pixel, or voxel for short.
These voxels all represent a small volume of the galaxy.
Since the voxels are perfectly aligned with the \hi data cube channels, 
calculating the brightness of each pixel in a channel can be done by 
performing the radiative transfer along the depth axis of the voxels 
associated with that pixel.

Each voxel is assigned coordinates $x$, $y$ and $z$.
These are initially aligned with the horizontal, vertical and depth axis.
We will subsequently bend and twist these coordinates to match the position 
and angle of the galaxy as seen in the cube. 
We emphasise however that these are \emph{internal} coordinates, the grid of 
voxels itself does not change and \emph{always} remains aligned with the 
\hi data cube.
The internal coordinates will represent the coordinate system of the galaxy 
in question.

\subsection{Generating the Galaxy Coordinate System}
We start  with the coordinates $x,y,z$ assigned to the grid of 
voxels\footnote{As an aid to understanding the coordinate system, we 
illustrate the steps these equations perform in Figure \ref{fig:cartoon}.}.
We first shift the coordinates, such that position (0,0,0) aligns with the 
centre of the model, i.e.
\begin{equation}
 \begin{pmatrix}x\\y\\z\end{pmatrix} =  \begin{pmatrix}x - \max(x)/2 \\ y - \max(y)/2\\ z - \max(z)/2\end{pmatrix}\,\,.\label{eqn:cartoon1}
 \end{equation}

When performing a fit, the actual galaxy is often not perfectly aligned with 
the centre of the \hi data cube.
It is thus necessary to shift the coordinates, based on parameters 
$\mathrm{shift}_x$ and $\mathrm{shift}_y$, 
\begin{equation}
 \begin{pmatrix}x\\y\end{pmatrix} =  \begin{pmatrix}x - \mathrm{shift}_x\\y - \mathrm{shift}_y\end{pmatrix}\,\,.\label{eqn:cartoon2}
\end{equation}

Next, the position angle $PA$ is applied. This is done using the coordinate 
transforms
\begin{equation}
 \begin{pmatrix}x\\y\end{pmatrix} =  \begin{pmatrix} x \cos(PA) - y \sin(PA) \\  x \sin(PA) + y \cos(PA)\end{pmatrix}\,\,.\label{eqn:cartoon3}
\end{equation}

Followed by the inclination $i$,
\begin{equation}
 \begin{pmatrix}x\\z\end{pmatrix} =  \begin{pmatrix} x \cos(i) + z \sin(i) \\  -x \sin(i) + z \cos(i)\end{pmatrix}\,\,.\label{eqn:cartoon4}
\end{equation}

We can now convert to the final cylindrical coordinates system that will be 
used for the galaxy,
\begin{equation}
 \begin{pmatrix}R\\ \theta\\z\end{pmatrix} =  \begin{pmatrix} \sqrt{x^2+y^2} \\  \arctan(y/x)\\ z \end{pmatrix}\,\,.\label{eqn:cartoon5}
\end{equation}

Warps are known to exist in many galaxies, yet the actual physics is not well 
understood.
The traditional way of dealing with warps is by applying a tilted-ring model.
This is however impossible in our case, as the tilted-ring model does not map 
all coordinates to a unique physical position. 
Some voxels would then be mapped to multiple values of $R$ simultaneously, 
which is undesirable.

We thus choose a different strategy to account for warps. 
Following \citet{Binney1987A}, we define a warp as a harmonic oscillation 
along the disc, in which the mid-plane $z=0$ is raised, or lowered, above the 
initial mid-plane, as we go around the galaxy following $\theta$.
The warp itself has a maximum amplitude which is described by super-function 
$z_\mathrm{warp}(R)$ (see Section \ref{sec:superfunctions}), which occurs at 
pitch angle $\theta_\mathrm{warp}$.
We do not account for radial changes in pitch angle, so $\theta_\mathrm{warp}$ 
is a single value. Then
\begin{equation}
 z = z + z_\mathrm{warp}(R) \times \cos(\theta + \theta_\mathrm{warp})\,\,.
\end{equation}

\subsection{Generating the Physics}
At this point, we have successfully generated the coordinates for the galaxy.
The next step is to calculate the actual physical properties for each voxel.
First, the line-of-sight velocity $v_\mathrm{obs}$ is calculated, based on 
rotation velocity $v_\mathrm{rot}$ and systemic velocity $v_\mathrm{sys}$. 
The rotation velocity $v_\mathrm{rot}$, along with the scale-height of the disc 
$z_0(r)$, velocity dispersion $\sigma(R)$ and face-on surface density 
$A_\textrm{\hi}(R)$ are super-functions, which are prepared by the user (see 
Section \ref{sec:superfunctions}). Then
\begin{align}
v &= v_\mathrm{rot}(R) - L \times \!\mid \!z\!\mid \label{eqn:lagginghalo}\,\,\,\,\,\,\,(v\geq0),\\
 v_\mathrm{obs} &= v \times \sin(\theta) \times \sin(i) + v_\mathrm{sys}\,\,.
\end{align}
Equation \ref{eqn:lagginghalo} offers the possibility to model lagging haloes.
The lag is expected to decrease linearly the observed rotation with value $L$ 
as the height $\mid \!z\! \mid$ above the plane increases.
The model automatically prevents $v<0$ and forces these to $v=0$.
When no halo lagging is modelled $L$ is set to zero.

The observed velocity dispersion $\sigma$ is calculated separately for the 
three (cylindrical) directions $\sigma_{z,\mathrm{obs}}$, $\sigma[B_{R,\mathrm{obs}}$ 
and $\sigma_{\theta,\mathrm{obs}}$, and then combined into a single observed 
velocity dispersion $\sigma_\mathrm{obs}$.
The $a_R, a_z, a_\theta$ components are scalars to account for a possible 
anisotropic velocity-dispersion tensor. \label{sec:tensordispersion}
There is currently no known case of anisotropy, so we set $a_R, a_z, a_\theta$ 
to one by default.
We also correct for instrument broadening using dispersion 
$\sigma_\textrm{inst}$, based on the channel-width $dv$. Then
\begin{align}
 \sigma_{z,\mathrm{obs}}      &= a_z \sigma(r) \times \sin(\theta) \times \sin(i)\,\,,   \\
 \sigma_{R,\mathrm{obs}}      &= a_R \sigma(r) \times \cos(i)                    \,\,,   \\
 \sigma_{\theta,\mathrm{obs}} &= a_\theta \sigma(r) \times \sin(i) \times \sin(\theta)\,\,,   \\
 \sigma_\textrm{inst}         &= dv / 2.355\,\,,\\
 \sigma          &= \sqrt{\sigma_{z,\mathrm{obs}}^2 + \sigma_{R,\mathrm{obs}}^2 +  \sigma_{\theta,\mathrm{obs}}^2 + \sigma_\textrm{inst}^2}\,\,.
\end{align}

The system takes the face-on surface density $A_\textrm{\hi}(R)$ and, using the 
flaring $z_0(R)$, converts it into a mid-plane density $\rho_\mathrm{m}(R)$. 
From this, the actual density $\rho$ for each voxel is calculated.
We treat the gas flaring as a Gaussian distribution. 
There exist alternative distributions, but the Gaussian has some advantages 
for the analysis of the hydrostatics, in Paper V \citep{Peters2015E}.
Based on \citet{Olling1995A}, the errors due to this are expected to be minor. 
Then
\begin{align}
 \rho_\mathrm{m} &= \frac{A_\hi(R)}{\sqrt{2 \pi z_0(R)^2}} \,\,,\\
 \rho &= \rho_\mathrm{m}  \exp\left[-\frac{z^2}{2 \,z_0(R)^2}\right]\,\,.\label{eqn:gaussiandisc}
\end{align}

The densities $\rho$ are all in units of \hi atoms per cm$^{3}$. 
In the final step, we convert to a local column density $N_\mathrm{HI}$ 
cm$^{-2}$ by multiplying with the physical length $ds$ that a single voxel 
represents along its depth axis.
\begin{equation}
 N_\mathrm{\hi} = \rho \,\, ds
\end{equation}

\subsection{Generating the \hi data Cube}
We have now prepared the 3D cube and it is time to perform a ray-trace through 
the voxels along the depth axis and generate the final \hi data cubes.
\textsc{Galactus} supports two integration modes: optically thin and 
self-absorption.

\subsubsection{Optically thin}
We assume that the \hi follows a Gaussian velocity distribution,
\begin{equation}
N_\mathrm{\hi} dv = \frac{ N_\mathrm{\hi}}{\sqrt{2\pi\sigma^2}}  \exp\left[-\frac{(v-v_\mathrm{obs})^2}{2\sigma^2}\right] \,\,.\label{eqn:phi_v}
\end{equation}

For a single channel in the \hi data cube, starting at velocity $v_1$ and with a 
channel width $dv$, the number of atoms $N$ in a voxel that are contributing 
to the emission in a single channel, can be calculated as
\begin{equation}
N = \int_{v_1}^{v_1+dv}{N_\mathrm{\hi} \,dv}\,\,.
\end{equation}

We solve this integral using the error function \emph{erf}, i.e.
\begin{equation}
N = \frac{N_\mathrm{\hi}}{2  dv} \left[ {\rm erf }\left(\frac{v_1 + dv - v_\mathrm{obs}}{\sqrt{2\pi\sigma^2}}\right) - {\rm erf }\left(\frac{v_1 - v_\mathrm{obs}}{\sqrt{2\pi\sigma^2}}\right)    \right] \label{eqn:Npervoxel}
\end{equation}

To calculate the optically thin surface-brightness of each voxel, we use the 
optically thin limit of the radiative transfer equation 
\citep[Equation 8.26]{Draine2011}, i.e.
\begin{equation}
 T_{B,\textrm{thin}} = \frac{N}{ 1.8127\times10^{18}\,dv}\,\,.\label{eqn:Tpervoxel}
\end{equation}

As there is no self-absorption, the intensity of a pixel can simply be 
calculated by taking the sum of $T_B$ in all voxels along the depth axis 
at the position of that pixel.

\subsubsection{Self-absorption}
When running in self-absorption mode, the calculation gets more complex.
We follow the method from the previous subsection up to equation 
\ref{eqn:Tpervoxel}, which gives us the emitted temperature in the velocity 
range of the channel per voxel for an optically thin model.
From this, we can then calculate the optical depth $\tau_\nu$ of that voxel 
as\footnote{Note that this is just a rewrite of Equation \ref{eqn:Tpervoxel} 
with Equation 8.11 in \citet{Draine2011}.},
\begin{equation}
\tau_\nu = T_{B,\textrm{thin}} / T_\mathrm{spin}\label{eqn:tauTT}\,\,.
\end{equation}

We assume that the galaxy has no internal continuum absorption, nor a 
background continuum source. 
This assumption is at least true for the galaxies in Paper I, although it 
does not hold for every galaxy.
To calculate the brightness of a pixel, we begin for the voxel at the back 
of the depth axis associated with that pixel, and calculate the radiative 
transfer equation with $T_\mathrm{bg}=0$ \citep[Equation 8.22]{Draine2011}.
\begin{equation}
T_{B,\textrm{self-absorbing}} = T_\mathrm{bg} \textrm{e}^{-\tau_\nu} + T_\mathrm{spin} \left(1 - \textrm{e}^{-\tau_\nu}\right)\label{eqn:self-absorption}
\end{equation}
For subsequent voxels $T_\textrm{bg}$ is the $T_B$ from the previous voxel.
We continue this for all voxels along the depth axis until the final value 
$T_B$, which is adopted as the brightness for that pixel.

In our own Galaxy, many lines of sight are optically thick within a few 
hundred parsec \citep{Allen2012A}. 
To ensure sufficient accuracy, it is therefore advised to choose the physical 
length of each voxel along the depth axis, such that it does not represent 
more than a hundred parsec.
This can be done using the accuracy parameter (see Section 
\ref{sec:supersampling}).

We note that the use of a constant spin temperature $T_\textrm{spin}$ and a 
uniform density $\rho$ per voxel serves only as an 
approximation.\label{sec:uniformdensity}
In reality the \hi in galaxies consists of gas at a range of spin 
temperatures, not to mention varying densities (e.g. clouds) on scales 
much smaller than a hundred parsec. 
We discuss this in more detail in Section \ref{sec:effectiveTspin}.

\subsection{Beam Smearing}
After the \hi data cube has been generated, we convolve the cube with the beam 
$\theta_\textrm{FWHM}$.
Note that we require the beam to be circular and the cells to be square.

\subsection{Simulating Noise}
It is possible to introduce artificial noise into a cube.
This can be done using the addition of the \texttt{-N} argument to the command 
line call to \textsc{Galactus}.
The noise $\sigma$ in Table 3 of Paper I was measured as 
the standard deviation per pixel.
In reality however, the noise is correlated between neighbouring pixels due to 
the beam smearing.
\textsc{Galactus} automatically corrects for this, such that the noise 
$\sigma$ is equal to that as measured from the standard deviation of the 
convolved field.

\subsection{Input}
\textsc{Galactus} can be used in two modes, stand-alone and as a software 
library.
To get the stand-alone version, the user can download the source code of the 
program from the dedicated website at sourceforge.net/p/galactus.
After compiling the required \textsc{C++} libraries, the program can then be 
run from the program folder using the command \texttt{./Galactus ini-file}. 
\textsc{Galactus} uses ini-files as its main source of information, and any
run of the program thus needs to be set up using them.
We provide a list of all the options in the online Appendix.

It is also possible to import \textsc{Galactus} into your own \textsc{Python} 
programs as a library.
All libraries required to run the program are contained in the 
\texttt{support/} module. 
It is beyond the scope of this paper to discuss this in detail 
and we refer to the documentation in the code itself for more details.
In short, to generate a basic model, a programmer would first need 
to import the \texttt{Galaxy} class from the \texttt{support.parameters} 
module, initialise a version of it, and then use the 
\texttt{parse\_ini(filename)} method to initialise all parameters 
in the \texttt{Galaxy} object.
The \hi data cube can then be generated by calling the 
\texttt{support.mainmodel.model()} method.
The source code has been extensively documented, and we invite 
developers to have a look and contribute to the code-base.

\subsubsection{Superfunctions}\label{sec:superfunctions}
While most variables in \textsc{Galactus} are single valued, the 
rotation curve and face-on surface density are examples of so-called 
super-functions.
Super-functions are a special class of functions that can behave 
differently based on the preference of the user.
In their 1 or 2 modes, (see Section \ref{tbl:Galactusinput}) they 
act as parametrised functions, while in the 3 and 4 modes they act 
as tabulated functions.
In tabulated mode, the user specifies the values of the function that 
the function has at positions $R$. 
During runtime, \textsc{Galactus} performs a linear interpolation 
over these points to calculate the value of intermediate radii.
We have experimented with alternative types of interpolation, such 
as spline interpolation.
While favourable from a theoretical point of view, in practise we found that 
during fitting a spline could start to show extreme spikes (both negative and 
positive), in an attempt to still reach all the user specified $R$ and $f(R)$ 
values. 
This made attaching boundary conditions very hard.
In the end, we have settled for linear interpolation between values.

In parametrised mode, the functions are simplified to
an analytical form that best fits their common profiles. 
For example, the velocity curve $v_\textrm{rot}$ has a parametrised form of 
$v_\textrm{rot}(R) = a + a / (R/b+1)$.
The advantage of parametrizing functions is that it limits the amount of 
free parameters, in this case only two. 
The form of each super-function is shown in Table \ref{tbl:Galactusinput}.
Another advantage to parametrised functions is that they are less sensitive 
to noise -- although at the expense of accuracy -- and give the ability 
for a very quick first estimate of the parameters.
We provide a function \texttt{all\_to\_spline}, 
which can convert parametrised curves to tabulated ones.

\subsection{Output}
\textsc{Galactus} is capable of producing a range of products, 
which can be selected in the parameter file.
The main output of the program is the generated \hi data cube.
There are however, more options. In this subsection, we will highlight the 
most important ones.

\subsubsection{Total, visible and hidden matter cube}
Since the brightness can depend on opacity, the generated \hi data cube no 
longer represents the total amount of atoms. 
The total matter cube gives the total amount of atoms in a velocity range, 
before any self-absorption is applied.
This starts with the result of Equation \ref{eqn:Npervoxel}, which is a cube 
with the amount of \hi per voxel, for a given channel. 
Rather than converting the data into temperatures with Equation 
\ref{eqn:Tpervoxel}, we instead sum the voxels along the depth axis 
and thus get the number of atoms.
This process is repeated for each channel in the \hi data cube.

In contrast, the visible matter cube uses the generated \hi data cube and 
converts it back into atoms, using the inverse of the optically thin 
approximation from Equation \ref{eqn:Tpervoxel}.
When running in optically thin mode, the visible and total matter cubes 
are equal.
The difference comes into play when running in self-absorption mode. 
This difference can be visualised with the hidden matter cube, which is 
a cube showing the difference between the total and visible matter cubes.

\subsubsection{Moment maps}
Besides channel maps, \textsc{Galactus} can also output various moment 
maps, specifically the zeroth, first, second and third moments.
These follow the same moment definitions as \textsc{Gipsy} \citep{GIPSY}.

\subsubsection{XV map and integrated profile}
\textsc{Galactus} is also capable of plotting the integrated XV (otherwise 
known as integrated PV) diagrams of the \hi data cube, by integrating the 
generated \hi data cube along the vertical axis.
By also integrating along the velocity axis, the program can generate an 
integrated profile.

\subsection{Fitting}
Fitting a galaxy model to a \hi data cube is a very complex problem, and great 
care needs to be taken in order to fit the data properly.
To give the user the best capabilities, we have included a number of fitting 
routines in the program.
Suppose one is fitting a well-resolved face-on galaxy with fixed velocity 
dispersion and no beam smearing. 
In this case, a line-of-sight corresponds to a single radius inside the galaxy.
This radius is sampled by many pixels as we look around the galaxy.
The parameter space is thus very limited, as each radius only has to fit a 
single face-on surface density and rotation curve.
A simple fitting routine, such as the Levenberg-Marquardt algorithm 
\citep{Press1992}, can quickly converge to the optimal solution.
We make use of the \textsc{Scipy} package \texttt{optimise.leastsq}.

When beam smearing is involved, the light from various radii starts to 
overlap, which causes the parameter space to get far more complex. 
Even worse, local optimal solutions start to appear.
For example, where the true face-on surface density at a radius might be 
$5\times10^{20}$ atoms/cm$^2$, the model could be happily converging towards 
$3\times10^{20}$ and $7\times10^{20}$ atoms/cm$^2$
in two adjacent radii, with the beam 
taking care of the smoothing back to $5\times10^{20}$ atoms/cm$^2$ on average.
The velocity dispersion also creates many local optima, especially in edge-on 
galaxies where the use of a very high velocity dispersion at a high radius 
$R$ can mask out the velocity dispersion at a slightly lower radius, thus 
convincing the program that low velocity dispersions at that radius are also 
acceptable. 

To find the global optimum, a more powerful algorithm is required.
We have extensively tested many algorithms.
Our preferred method is the \textsc{PSwarm} algorithm, which is a combination 
of pattern search and particle swarm 
\citep{Vaz07,Vaz09,Thi12}\footnote{Website: www.norg.uminho.pt/aivaz/pswarm/} 
and is used through the \textsc{OpenOpt} 
framework\footnote{Available at openopt.org/.}.
Solving for a global optimization problem requires a far more thorough search 
of the parameter space and thus requires much more time to converge, compared 
to the Levenberg-Marquardt algorithm.
We have also tested various implementations of genetic algorithms, but found 
that the random nature of the improvements led to unacceptably long 
convergence times.

We also find that when fitting for the velocity dispersion, it is often 
misused by an algorithm to 'average over' local structure in the XV 
diagram, by using a very high velocity dispersion. 
This can happen in particular when warps and self-absorption are 
present, but are not allowed for in the modelling. 
In those cases, it is not even possible to model the XV diagram 
accurately. The algorithm will then start to misuse the velocity dispersion.

To estimate the $\chi^2$ cost of some parameter set $p$ on the model 
$T_\textrm{model}(p)$, we use least-squares deviations.
We show this in Equation \ref{eqn:lossfunction}, where we use the standard deviation of the noise $\sigma$ from the observed \hi data cube $T_\textrm{observation}$.
Index $i$ runs over all pixels in the \hi data cube,
\begin{equation}
 \chi^2(p) = \sum_{i=0}^{N_\textrm{pixels}}{\frac{\left[T_\textrm{model}(i;p) - T_\textrm{observation}(i)\right]^2}{\sigma^2}}\,\,.   \label{eqn:lossfunction}      
\end{equation}

Lastly, we implement a Monte-Carlo Markov-Chain (MCMC) version of the model. 
MCMC is a different approach to model optimisation. \label{sec:Galactus-MCMC}
Rather than look for a single parameter set that represents the global optimal 
solution, MCMC explores the parameter space and draws samples of parameter 
sets from the region where, based on the probability distribution, the optimum 
is expected.
We make use of the \textsc{emcee} library \citep{emcee}, which implements the 
affine-invariant ensemble sampler proposed by \citep{emcee2}.
Following the cost function $\chi^2$, given in Equation \ref{eqn:lossfunction}, 
the log-likelihood $\mathcal{L}$ function for a particular parameter-set $p$ 
is calculated as
\begin{align}
 \mathcal{L}(p) &= \log{\left(-\frac{\chi^2(p)}{2}\right)}\,\,,\\
             &= \log{\sum_{i=0}^{N_\textrm{pixels}}{-\frac{\left[T_\textrm{model}(i;p) - T_\textrm{observation}(i)\right]^2}{2\sigma^2}} }\,\,.\label{eqn:laplacedistribution}
\end{align}

\textsc{emcee} uses multiple MCMC samplers in parallel, each of which generates 
a chain of samples.
For the number of samplers we choose double the amount of free parameters, 
which for tabulated fitting implies 150-300 chains.
After a sufficiently long burn-in period, we collect a large group of $N$ 
samples from the collection of chains.
For each free parameter that is being fitted, 
the group thus has $N$ values 
that this parameter would most likely have. 
These values thus form a distribution.
In this and subsequent papers, we make use of the central 68, 95 and 99.7\% 
of the likelihood distribution to denote the errors.
Note that if the distribution in a parameter follows a Gaussian distribution, 
this would be equal to the one, two and three sigma deviations of that Gaussian.

\section{Notes}\label{sec:notes}
\subsection{Masking the Data}
\textsc{Galactus} can deal with masked pixels.
Pixels that are masked in the input \hi data cube, specified with the \
texttt{image} parameter (see also Section \ref{tbl:Galactusinput}), are 
treated as pixels with a value of zero.
\textsc{Galactus} will still render the corresponding pixels in the 
output model, as the user could still be interested in the model at 
these pixels, but has masked them on purpose for some reason.

The \texttt{mask} parameter can be used to specify which pixels 
\textsc{Galactus} does not need to render. 
The parameter needs to point to a \hi data cube, which can be the 
same as the one used in the \texttt{image} parameter. 
All pixels in the channel maps that make up the \hi data-cube that 
have value NaN\footnote{NaN stands for Not A Number. By default, 
\textsc{Miriad} masking gives pixels this value.} are selected 
as the masked pixels.
\textsc{Galactus} calculates which voxels contribute to which pixels. 
If a voxel only contributes to masked pixels, it is removed from the 
calculation. 
This can lead to drastic performance increases.

The last type of mask is the \texttt{boundaries} mask. 
This can be used to specify in which channel maps the line-of-nodes 
velocity $v_\textrm{rot}(R)$ is expected to lie.
It is then possible to mask everything except the outer envelope of a XV 
diagram. 
The fitting algorithms are then constrained in the  $v_\textrm{rot}(R)$ 
parameter.

\subsection{Concerning the Resolution}\label{sec:supersampling}
By default, \textsc{Galactus} traces only one ray per pixel.
The initial cube has voxels that are one pixel wide on all sides. 
Thus, if the output cube has pixels with a (projected) size of one kpc, 
the voxels are one kpc on each side as well, and the ray is thus sampled 
every one kpc as it crosses the galaxy.
For a galaxy that has a diameter of 20 kpc, the rays will thus at most only 
sample 20 positions inside the galaxy.
This can lead to a jagged looking result.

To provide a better representation of the galaxy, we have introduced two 
additional user options.
The first is the \texttt{accuracy} level, which controls the number of 
samples along the ray. 
It works as a scalar on the original number of samples along the depth 
axis, such that if one originally measured 20 samples, an accuracy level 
of four would lead to 80 samples. 
Following the above example, these would thus be separated 250 pc apart 
along the ray.

The second option is \emph{super-sampling}, a technique common in the 3D 
graphics industry. 
In super-sampling, each output pixel is represented by multiple rays at 
slightly different positions inside the pixel, with the result being the 
average of all rays.
For example, suppose the original ray runs a trace centred on the middle 
of the pixel, thus at position 1/2. 
Then with super-sampling, we have two rays, both at 1/3 and 2/3. 
Suppose that a significant change would beyond position 1/2. The normal 
model would not represent this, while super-sampled pixel would.
Super-sampling works in both the horizontal and vertical directions.
A super-sampling of factor two means each pixel is sampled by four rays.

An increased accuracy thus increases the number of voxels along the depth axis.
An increase in super-sampling decreases the size of an individual voxel 
along the horizontal and vertical axis and thus increases the number of 
voxels quadratically.
The increases in accuracy and super-sampling are thus increases in the 
number of voxels that need to be calculated, which come with a performance 
penalty.
It is thus important to realise that there is a trade-off between the 
desired accuracy and the time available.

\begin{figure}
 \resizebox{0.48\textwidth}{!}{\includegraphics{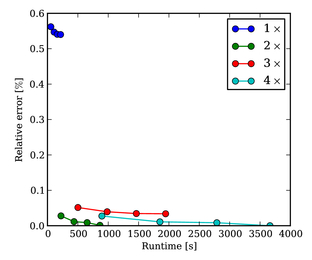}}
 \caption[Resolution trade-off in Galactus]{The accuracy versus time trade-off visualized. The various tracks show the various super-sampling modes, while the dots along each track represent the accuracy level used. Along each track, the top left point is always accuracy level of one, progressing to an accuracy level of four, at the right-hand side. Overall the rule holds that the closer to (0,0), the better.}
 \label{fig:AA}
\end{figure}

In Figure \ref{fig:AA}, we visualise this trade-off.
Shown are the timing results of an edge-on model, at multiple combinations 
of super-sampling and accuracy levels, with their relative error 
compared to a very high-resolution model with an accuracy of five and a 
super-sampling of five, which in this case implies a model with voxels 
that each represent 27 pc on all side. 
In a perfect world, the accuracy and super-sampling could be set to 
$\infty$ and the results would be perfect.
In practice however, due to time constraints, there is no single right setting, 
 it is thus important to think about the problem and decide what the 
best setting should be.
For example, in a face-on galaxy one quickly runs the risk of 'overshooting' 
the disc, 
 if there were not enough samples along the ray, and a higher accuracy would 
then be desirable.
In an edge-on, overshooting the disc is less important, but one might want to 
super-sample 
 the galaxy as to model the vertical structure more accurately.
 
A super-sampling of two incorporates a special feature, which positions the 
four rays in an optimal distribution.
In this way, both the horizontal and vertical axis of the pixel in question 
are traced at four unique positions, rather than the default of two positions 
per axis. 
The effect of this can be seen in Figure \ref{fig:AA}, where the results 
from a super-sampling of 2 are nearly equal to the far more computationally 
expensive super-sampling of four (i.e. 16 rays).
A super-sampling of two outperforms a super-sampling of three.
Figure \ref{fig:AA} also demonstrates that increasing the accuracy does not 
drastically improve the model.
 We recommend that the accuracy be chosen such that the length of the voxel 
along the depth axis is at most 100 parsec, such that the radiative transfer 
equation is calculated with sufficient precision.

\begin{figure*}
 \includegraphics[width=0.38\textwidth]{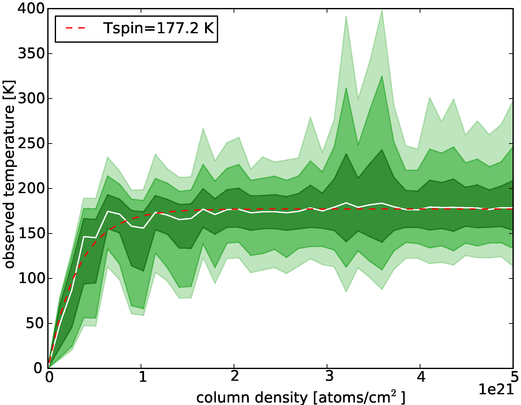}
 \includegraphics[width=0.38\textwidth]{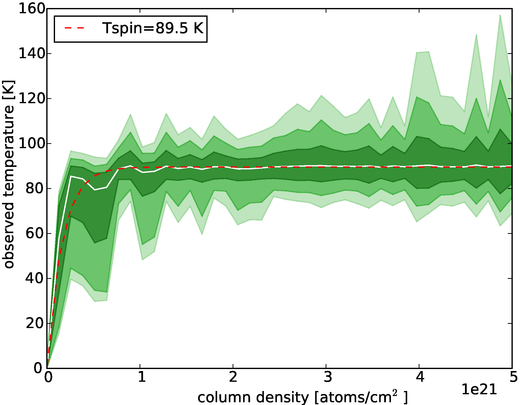}
 \caption[Effective spin temperature assuming $p_\textrm{CNM}=0.8$]{Effective spin temperature assuming $40\%$ (left panel) or $80\%$ (right panel) of the gas is in the CNM. Colour bars represent $68.5\%$, $95.45\%$ and $99.73\%$ of the distribution at each point. In white the $50\%$ (median) of the sample is shown. The red dashed line represents the fitted effective spin temperature.}
 \label{fig:effectivespin8}\label{fig:effectivespin4}
\end{figure*}

\begin{figure*}
 \includegraphics[width=0.80\textwidth]{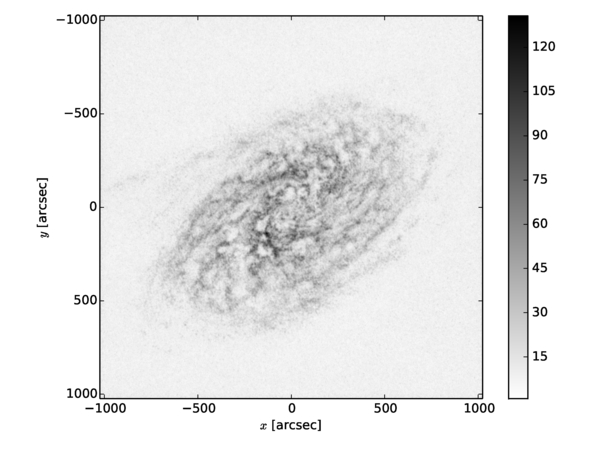}
 \caption{Maximum surface brightness map for NGC\,2403. The values for the bar on the right is brightness temperature in Kelvin.}
 \label{fig:NGC2403-maxtempmap}
\end{figure*}

\section{The Effective Spin Temperature}\label{sec:effectiveTspin}
In a real galaxy, the \hi never has a single spin temperature.
The gas is balanced between the phases of the CNM, which has a median spin 
temperature of 80\,K 
 and the WNM, with temperatures between 6000 and 10000\,K.
The fraction of the mass contained in either state remains unclear, 
 but it was estimated that $40\%$ of the neutral hydrogen mass of the Galaxy 
is in the CNM \citep{Draine2011}.

With such a wide range in temperatures, how reliable is the use of a single 
effective spin temperature?
To test this, we have run a range of Monte Carlo simulations.
For a fixed CNM fraction $p$, we calculate the effective spin temperature over 
a range of column densities.
For column density $N_\textrm{\hi}$, we divide it up in $N_\textrm{clouds}$ equal 
sized bins, each of which represents a cloud of gas with a column density 
$N_\textrm{cloud}$.
The number of clouds is randomly picked between 100 and 1000, to represent the 
variation in crossing a real galaxy.
Based on fraction $p$, we randomly choose if a cloud is part of either the CNM 
or the WNM.
If it is a CNM cloud, it is assigned a random spin temperature 
$T_\textrm{spin}$, chosen from a uniform distribution between 50 and 100\,K.
Otherwise, it is assigned a $T_\textrm{spin}$ chosen from a distribution 
between 6000 and 8000\,K.
We then perform a line-of-sight integration, starting  with background 
temperature $T_\textrm{bg}=0$\,K,  using Equations \ref{eqn:Tpervoxel}, 
\ref{eqn:tauTT} and \ref{eqn:self-absorption}.

Taking 1000 samples at each column density $N_\hi$, we get a distribution 
of observed temperatures $T_B$.
We show the results for two mass fractions in Figures \ref{fig:effectivespin4}a 
and \ref{fig:effectivespin8}b.
We fit a single effective $T_\textrm{spin}$ to the median of the distributions, 
using
\begin{equation}
T = T_\textrm{spin}\left(1 - e^{-\tau}\right)\,\,.
\end{equation}
The results are over-plotted in red in Figures \ref{fig:effectivespin4}a and 
\ref{fig:effectivespin8}b.
For the $40\%$ fraction, we find an effective spin temperature of 177.2\,K, 
while for the $80\%$ fraction we find 89.5\,K.
Clearly, the balance between the two populations is important.
More importantly, in both cases we see that the median spin temperature 
levels out, and a single effective spin temperature gives a decent fit 
to the median one.
Provided the right value is chosen, it is thus possible to use a single 
effective spin temperature when modelling a galaxy.
What the exact value for this effective spin temperature is, is hard to determine 
and is beyond the scope of this work.
During fitting, we shall use a spin temperature of 100\,K, 
a value in good agreement with the figures in Paper I.

\begin{figure}
 \includegraphics[width=.48\textwidth]{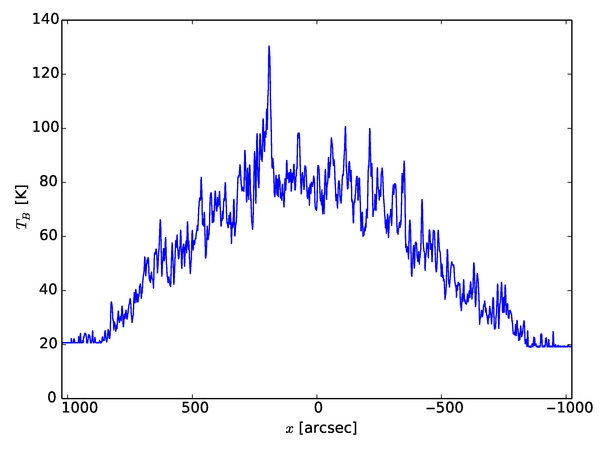}
 \caption{Radial maximum surface brightness of NGC\,2403}
 \label{fig:NGC2403-maxtempalongr}
\end{figure}

\section{Face-on Galaxy NGC~2403}\label{sec:NGC2403}
In the following sections we test the program for a number of cases. 
To begin with, we wish to test our fitting routines on a face-on galaxy to 
confirm the accuracy of the rotation curve and surface density modelling.
For this purpose, we have chosen the galaxy {NGC\,2403}. 
Like the galaxies in our sample it is a nearby, late-type Scd galaxy 
\citep{Tully2000A}. \hi observations of high spatial and velocity
resolution are available
for it. Even though the rotation velocity is slight higher (130 versus on 
average 90\kms) it is suitable for
comparing to our sample of edge-on, late-type galaxies (Paper I). 
Using the tip of the red giant branch, \citet{Dalcanton2009A} estimated the 
distance at 3.2\,Mpc.
The \hi kinematics have have been studied in detail by \citet{Sicking1997A}, 
\citet{Fraternali2001A,Fraternali2002A} and \citet{DeBlok2008A}. 
The galaxy is part of the public-data release from The \hi Nearby Galaxy 
Survey (THINGS), which was based on B-, C- and D- configuration Very Large 
Array (VLA) observations\footnote{The public data from the THINGS survey is 
available at www.mpia-hd.mpg.de/THINGS/Data.html.} \citep{Walter2008A}.
The THINGS observations were previously studied in detail by 
\citet{DeBlok2008A}, who measured a systemic velocity $v_\mathrm{sys}$ of 
132.8\,km\,s$^{-1}$ and an inclination $i$ of $62.9^\circ$.

We test \textsc{Galactus} using the naturally weighted \hi data cube from the 
THINGS website. 
The original cube has a size of $2048\times2048$ pixels and 61 channels, 
which is too large for \textsc{Galactus} to handle.
\textsc{Miriad} task \texttt{smooth} was used to create a circular beam 
with a full-width half-maximum of 12.4\, arcseconds.
We used the task \texttt{regrid} to downsize the resolution from 1.0 to 
8.8\,arcsec/pixel, resulting in somewhat less than 2 pixels/beam. 
The initial beam was $8.75\times7.65$ arcsec, with a position angle of 
25.2 degrees.
We used the task \textsc{regrid} to downsize the sampling of the data 
from 1.0 to 8.8 arcsec/pixel, resulting in somewhat less than 2 pixels/beam. 
We have chosen this sampling as this corresponds to roughly 100 parsec per 
pixel.
As each pixel is supersampled 
by 4 rays, the area of the FWHM of the beam will be modeled by of order 
20 rays, 
which is sufficient to prevent undersampling. While an even higher resolution 
would have been desirable, it was found to be computationally infeasable, as 
the required memory and computation time increase with the power 3.
Using task \texttt{cgcurs}, a mask was drawn by hand around all regions 
containing flux.
With task \texttt{imsub}, we used this mask to create a final, masked cube 
that only contains parts of the cube where flux was detected.
The final cube has a size of $232\times228$ pixels and 55 channels at 
5.169\,km\,s$^{-1}$ resolution.

In Figure \ref{fig:NGC2403-maxtempmap}, the maximum surface brightness has 
been calculated from the original THINGS image.
Here we have converted the intensity $I$ from Jy/beam to Kelvin using 
Equation 4 of Paper I:
\begin{equation}
 T_B = \frac{\lambda^2 S}{2 k_B \Omega} = \frac{606000 S}{\theta^2}\,\,.\label{eqn:JytoK}.
\end{equation}
A clear spiral structure is visible in the figure.
Most of the brightness hovers between 45 and 60\,K and appears as spiral arm 
structure, with local regions peaking at over 120\,K.
We rotated the galaxy such that the major axis is lined up with
the horizontal x-axis.
If we only look at the maximum brightness along the horizontal axis, such that 
each position reflects the maximum brightness independent of location along 
vertical axis, we get the result as seen in Figure 
\ref{fig:NGC2403-maxtempalongr}.
This produces a result similar to the maxima along the major axis in edge-on 
galaxies as seen in Paper I.
There we found a plateau around 80\,Kelvin in all edge-on galaxies for a 
significant part of their discs. 
There is no clear plateau in Figure \ref{fig:NGC2403-maxtempalongr}.
The wings of this profile are much extended, running continuously towards 
the background. 
This is different compared to our edge-on galaxies, where we observed very 
sharp wings.
In Paper I, we argued that the plateau in the profile was due to 
self-absorption in long lines-of-sight.
The shorter lines-of-sight in {NGC\,2403} would therefore not be 
expected to exhibit (strong) signs of self-absorption.
Only in the dense inner region do the densities still reach levels 
high enough for self-absorption to significantly affect them, as 
exhibited by the peak around 80\,K.
We thus argue that with the exception of the central part, most of 
the galaxy is not affected by significant self-absorption.
We note  in general, that observations of self-absorption can only 
occur when the beam has sufficient spatial resolution. Otherwise, 
the effect will wash out due to smearing.

We fit {NGC\,2403} using both an optically thin model and a self-absorption 
model. 
For the self-absorption model, we use a spin temperature $T_\mathrm{spin}$ 
of 100\,K.
We fit the galaxy using a constant velocity dispersion of 10\,km/s.
We constrain the disc to a thickness of 700\,pc in the inner part and 
flaring out to 1\,kpc at 1400'' ($23$\,kpc).
Warps or lagging haloes are beyond the scope of the current test.
A double-pass strategy is used to fit the data. 
We first use the parametrised functions for the face-on surface 
density, flaring and rotation curve. 
This pass is fitted using the Levenberg-Marquardt algorithm.
After that, we interpolate over the fitted functions and tabulate 
them into 37 chunks with a separation of 38''.
We then run a second pass.
As the galaxy is face-on and very well resolved there is no strong 
danger of running into local minima, and we thus again use the quick
 Levenberg-Marquardt algorithm.

\begin{figure*}
     \resizebox{0.48\textwidth}{!}{\includegraphics{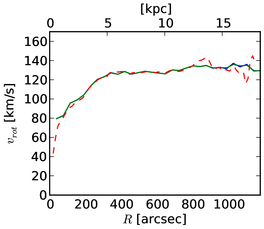}}
     \resizebox{0.48\textwidth}{!}{\includegraphics{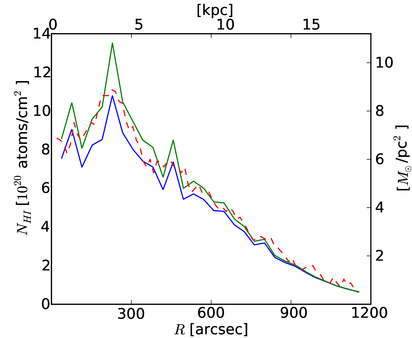}}
     \caption[Kinematic and mass models for NGC\,2403]{Kinematic and mass models for {NGC\,2403}. The blue profiles represent the optically thin model. The green profiles are the self-absorption fit. The dashed red profiles represent the data from \citet{Sicking1997A}.}
     \label{fig:NGC2403-fits}
\end{figure*}

The results for this fitting are shown in Figure \ref{fig:NGC2403-fits}.
The rotation curves of the optically thin and self-absorbing fits are 
effectively identical.
It is reassuring to find that in both cases the same rotation curve 
is found, proving that the observed $v_\textrm{rot}$ in face-on 
galaxies is not affected by self-absorption.
The profile also matches up well with the profile from \citet{Sicking1997A}.
Only in the outer radii do we suffer more from signal to noise issues.
Given the detailed tilted ring and warp fitted by \citet{Sicking1997A}, which 
we are ignoring, this is expected. 

Now consider the mass models.
Unsurprisingly, the optically thin profile is clearly lower than the 
optically thick profile.
The first has a total mass of $2.9\times10^9$\,M$_\odot$. 
The self-absorbing mass totals at $3.2 \times 10^9$\,M$_\odot$, of which 
$\sim10\%$ is hidden by the self-absorption.
The overall shape of the profiles agree well, with all local features 
visible in both profiles.
In comparison, \citet{Sicking1997A} found slightly more mass than our 
optically thin model.
They report a total mass of $3.27\times10^9$ solar masses for {NGC\,2403}.

\section{The Maximum Surface Brightness in an Edge-on Galaxy}\label{sec:NGC2403edgeon}
In Section 3 of Paper I, we presented a simple toy model to demonstrate 
that a plateau of constant maximum surface brightness is an indication of 
self-absorption.
Using \textsc{Galactus}, we can now extend this toy model to a more 
physical basis.
We project the previous fit of {NGC\,2403}  to a full edge-on orientation.
We use the self-absorption fit as the face-on surface-density and 
rotation curves as input parameters (see Figure \ref{fig:NGC2403-fits}).
We will compare between an optically thin model and a self-absorption 
run at a spin temperature of 100\,K.

\begin{figure}
 \includegraphics[width=0.48\textwidth]{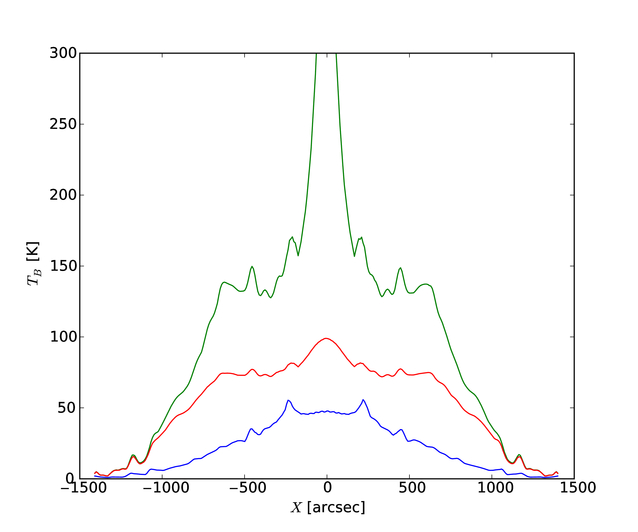}
 \caption[Theoretical maximum surface brightness temperatures for an edge-on galaxy]{Maximum surface brightness along the major axis based on modelling for {NGC\,2403}. 
 The lower blue line is the galaxy at its normal inclination. 
 The top green line is the optically thin case as seen edge-on. 
 The middle red line is the self-absorption case with a spin temperature of 100 K.}
 \label{fig:model-maxT}
\end{figure}

The results for this test are shown in Figure \ref{fig:model-maxT}. 
The model for the galaxy at its original inclination peaks at roughly 
40\,K (blue line). 
This is lower than the observed maximum brightness maps as seen in 
Figure \ref{fig:NGC2403-maxtempalongr}.
This is expected, as the model cannot recreate the small-scale 
bright regions seen in the actual observation.
When we rotate the galaxy to full edge-on,  we see that the long 
lines-of-sight carry the integration well above 300\,K (green line). 
Especially near the inner part at $x=0$ the model keeps rising.
As we have argued in Paper I, this is never observed. 
Comparing this to the self-absorption model at $T_\mathrm{spin}=100$~K, 
we see that the maximum surface brightness shows a clear plateau 
around 80 Kelvin with sharp edges.
Comparing this to the galaxy observations in Paper I, the self-absorption 
clearly matches the data better than the optically thin model.
Compared to the toy model from Section 3 of Paper I, the result is 
less dramatic than in Figure 2 in that paper, where we applied a 
far higher face-on column density distribution than used here.

\section{Effect of Inclination on the Visible Mass}\label{sec:effectofinclination}
Having established that an edge-on galaxy can easily hide a considerable 
fraction of its neutral hydrogen, we wish to test the relation between 
inclination and self-absorption.
We test the effect of self-absorption on galaxies in the range of 
inclinations from 60 to 90 degrees.
We have selected $T_\mathrm{spin}=100$\,K and $T_\mathrm{spin}=150$\,K 
as the spin temperatures we wish to model, and will produce an 
optically thin model at each inclination to compare.
We use the self-absorption results of {NGC\,2403} from Section 
\ref{sec:NGC2403} as the basis for all three models.

\begin{figure}
 \includegraphics[width=0.48\textwidth]{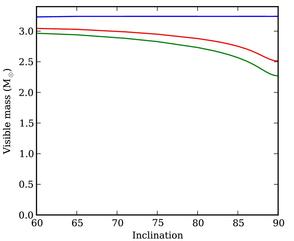}
 \caption[Visible \hi mass as function of inclination]{Visible \hi in an edge-on model based on {NGC\,2403}. 
 The top blue line is the galaxy as if there is no self-absorption present.
 The bottom green line has self-absorption at $T_\mathrm{spin}=100$\,K.
 The middle red line is the self-absorption at $T_\mathrm{spin}=150$\,K.}
 \label{fig:effect-of-inclination}
\end{figure}

The results for this fit are shown in Figure \ref{fig:effect-of-inclination}.
It is clear that in both cases self-absorption is present at all inclinations.
However, beyond an inclination of about 82 degrees, the effects starts 
to increase drastically.
Similar to Paper I, we find that for a spin temperature of 100\,K that 
$30\%$ of the \hi remains undetected.
The milder case of a spin temperature of 150\,K still has a significant 
absorption, with $\sim23\%$ of the \hi going unseen.

A natural prediction from Figure \ref{fig:effect-of-inclination} would 
be that in a statistical study the edge-on galaxies would have a lower 
average \hi mass than the face-on to moderately inclined galaxies.
We have tried to investigate this by comparing the K-band magnitude 
(which is expected not to be much affected by inclination effects) with 
the apparent \hi mass, using a combination of literature surveys. 
Unfortunately, we were unable to confirm or disprove the results, as 
there were too few edge-on galaxies in the available surveys with 
exact inclination measured.
It is beyond the scope of this project to investigate this 
in more detail.

\section{The Baryonic Tully-Fisher Relation}\label{sec:TullyFisher}

\begin{figure}
 \includegraphics[width=0.48\textwidth]{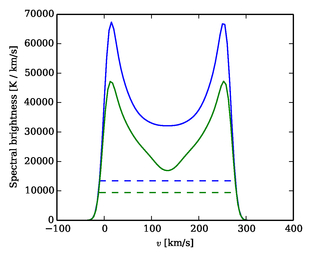}
 \caption[Tully-Fisher with self-absorption]{Impact on the width $w_{20}$ as measured from the spectra from an edge-on galaxy affected by in an optically thin (top, blue) and self-absorption with $T_\textrm{spin} =100$~K (bottom, green) case.}
 \label{fig:TF}
\end{figure}

Having established that edge-on galaxies can be affected by significant \hi 
self-absorption, what is the effect on the baryonic Tully-Fisher (TF) 
relationship? 
The TF relation relates the dynamical mass of a galaxy to its 
luminosity and provides a test of theories of galaxy formation and 
evolution \citep{Verheijen1997A,Bell2001A}.
We use the two edge-on models from Section \ref{sec:NGC2403edgeon} 
as a basis for this analysis.
The integrated spectra are shown in Figure \ref{fig:TF}.
The T-F relation is based on the width $w_{20}$ of the profile at 
20\% of the maximum in the integrated spectra \citep{Tully1977A}.
Clearly, this height varies drastically between the two cases, yet 
the difference between the two $w_{20}$ measurements is minimal.
This is due to the sharp drop-off that both profiles show at the 
extremes, where self-absorption is less important.
For the optically thin case a value of 285.0 km/s is found and 
for the self-absorption case 291.0\,km/s, an increase of only $2\%$.
We thus conclude the $w_{20}$ parameter is not affected
significantly by self-absorption.

The baryonic TF relation is however also based on the 
total baryonic mass $M_\textrm{bar}$, and as Figure \ref{fig:TF} 
clearly illustrates, self-absorption can significantly affect that property.
The TF-relation has not been well set, with various authors 
finding slopes between 
$M_\textrm{bar}\propto V^3$ and $M_\textrm{bar}\propto V^4$ (\citet{kf11} and 
references therein).
Although beyond the scope of this project, it will be interesting to see 
if the intrinsic scatter of the T-F relation decreases when 
self-absorption of the \hi is taken into account.

\section*{Acknowledgments}
SPCP is grateful to the Space Telescope Science Institute, Baltimore, USA, the 
Research School for Astronomy and Astrophysics, Australian National University, 
Canberra, Australia, and the Instituto de Astrofisica de Canarias, La Laguna, 
Tenerife, Spain, for hospitality and support during  short and extended
working visits in the course of his PhD thesis research. He thanks
Roelof de Jong and Ron Allen for help and support during an earlier 
period as visiting student at Johns Hopkins University and 
the Physics and Astronomy Department, Krieger School of Arts and Sciences 
for this appointment.

PCK thanks the directors of these same institutions and his local hosts
Ron Allen, Ken Freeman and Johan Knapen for hospitality and support
during many work visits over the years, of which most were 
directly or indirectly related to the research presented in this series op 
papers.

Work visits by SPCP and PCK have been supported by an annual grant 
from the Faculty of Mathematics and Natural Sciences of 
the University of Groningen to PCK accompanying of his distinguished Jacobus 
C. Kapteyn professorhip and by the Leids Kerkhoven-Bosscha Fonds. PCK's work
visits were also supported by an annual grant from the Area  of Exact 
Sciences of the Netherlands Organisation for Scientific Research (NWO) in 
compensation for his membership of its Board.


\bibliography{refsII}
\bibliographystyle{mn2e}


\appendix

\section{Overview of Input File Layout and Settings}\label{tbl:Galactusinput}
\textsc{Galactus} uses ini-files as input. 
Each ini-file is divided into various segments that detail the various parts 
of the program.
\bigskip

\begin{description}
  \item[\texttt[cube]] \hfill \\
  \begin{description} 
  \item[\texttt{image}] A list of the filenames to which we are fitting. \\Example: \texttt{["galaxies/UGC7321.fits.gz"]}.
  \item[\texttt{name}] The name of the object. All output files will have this as a base-name. Example: \texttt{UGC7321}.
  \item[\texttt{comparison}] Used for comparing to other ini files. The parameters from this file are also over-plotted in the output profiles. Example: \texttt{galaxies/UGC7321.ini}.
  \item[\texttt{mask}] Use a blanking mask in which all pixels of value \texttt{NaN} are ignored. This will speed up the program and changes which pixels are used while fitting. Example: \texttt{galaxies/UGC7321.fits.gz}
  \item[\texttt{boundaries}] This file controls which boundaries the rotation curve fitting has to oblige to. Setting parts of this cube to \texttt{nan} will ensure that the $v_\textrm{rot}$ will avoid that position. \\Example: \texttt{galaxies/UGC7321.fits.gz}
  \item[\texttt{cell}] Size of each pixel in arcsec. \textsc{Galactus} requires the pixels to be square. Example: \texttt{3.26}. 
  \item[\texttt{v\_cube\_min}] Velocity of the first channel in km/s. Example: \texttt{1565.0}. 
  \item[\texttt{v\_cube\_N}] Number of channels in the cube. Example: \texttt{75}.
  \item[\texttt{v\_cube\_delta}] Channel width and spacing, in km/s. Example: \texttt{3.298}
  \item[\texttt{width}] Length of the horizontal axis of the cube, in pixels. \\Example: \texttt{80}.
  \item[\texttt{height}] Length of the vertical axis of the cube, in pixels. Example: \texttt{25}.
  \item[\texttt{halforwhole}] Lets the program know if we are modelling the full cube or one side. This sets the length of the depth axis. Example: \texttt{whole}.
  \end{description}
\bigskip

  \item[\texttt{[physical]}] \hfill \\
  \begin{description}
    \item[\texttt{shift\_x}] Offset from the centre of the galaxy, compared to the centre of the radio channel, along the horizontal axis, in pixels. Example: \texttt{0}.
    \item[\texttt{shift\_y}] Offset from the centre of the galaxy, compared to the centre of the radio channel, along the vertical axis, in pixels. Example: \texttt{-0.2}.
    \item[\texttt{shift\_x\_free}]. Do we have to fit for the \texttt{shift\_x} parameter?\\ Example: \texttt{yes}.
    \item[\texttt{shift\_y\_free}]. Do we have to fit for the \texttt{shift\_y} parameter?\\Example: \texttt{no}.
    \item[\texttt{v\_sys}] Systemic velocity of the galaxy, $v_\textrm{sys}$, in km/s. Example: \texttt{409.3}.
    \item[\texttt{v\_sys\_free}] Fit for the systemic velocity? Example: \texttt{yes}.
    \item[\texttt{v\_sys\_bounds}] Tuple with the boundaries between which $v_\textrm{sys}$ should be, in km/s. Example: \texttt{(390.0, 420.0)}.
    \item[\texttt{distance}] Distance $D$ of the galaxy in Mpc. Example: \texttt{10.0}.
    \item[\texttt{distance\_free}] Fit for the distance of the galaxy? Note that this is often not a recommended strategy. Example: \texttt{no}.
    \item[\texttt{distance\_bounds}] Tuple with the boundaries between which the distance $D$ should be found, in Mpc. Example: \texttt{(9.6, 10.4)}
    \item[\texttt{T\_spin}] Spin temperature of the galaxy. Currently only global spin temperatures are supported. Select \texttt{None} for optically thin. Otherwise, the user can give a positive value, which will be taken as the spin temperature in Kelvin.
    \item[\texttt{rotationcurve\_function}] Which rotation-curve super-function should be used? Select \texttt{1} or \texttt{2} when using function $v_\textrm{rot}(R) = a + a / (R/b + 1)$, where \texttt{1} select no-fitting and \texttt{2} selects fitting. Selecting \texttt{3} or \texttt{4} sets for tabulated interpolation, with \texttt{3} for no fitting or \texttt{4} for fitting.
    \item[\texttt{rotationcurve\_x}] Either an empty list (e.g. \texttt{[]}) when \\ \texttt{rotationcurve\_function} is 1 or 2. Otherwise, a list of positions (expressed in pixels) along $R$, example: \texttt{[0,1,2,3,4,5]}. When 0 is omitted, the program assumes $R=0$ will give a $v_\textrm{rot}(R=0)=0$. 
    \item[\texttt{rotationcurve\_y}] Either a two-values  list (e.g. \texttt{[$A$,$B$]}) when \\ \texttt{rotationcurve\_function} is 1 or 2. Otherwise a list of rotation curve strengths $v_\textrm{rot}$ (expressed in km/s) along $R$, example: \texttt{[10, 20, 40, 80, 90]}.
    
    \item[\texttt{surfacedensity\_function}] Which face-on-surface-density super-function should be used? Select \texttt{1} or \texttt{2} when using function $A_\hi(R) = |a| + b R + c R^2$, where \texttt{1} select no-fitting and \texttt{2} selects fitting. Selecting \texttt{3} or \texttt{4} sets for tabulated interpolation, with \texttt{3} for no fitting or \texttt{4} for fitting.
    \item[\texttt{surfacedensity\_x}] Either an empty list (e.g. \texttt{[]}) when \\\texttt{surfacedensity\_function} is 1 or 2. Otherwise a list of positions (expressed in pixels) along $R$, example: \texttt{[0,1,2,3,4,5]}. 
    \item[\texttt{surfacedensity\_y}] Either a three-value  list (e.g. \texttt{[$a$,$b$,$c$]}) when \\\texttt{surfacedensity\_function} is 1 or 2. Otherwise a list of face-on surface densities $A_\hi$ (expressed in $10^{20}$ atoms/cm$^2$) along $R$, example: \texttt{[5,4,2,1,0]}.

    \item[\texttt{flaring\_function}] Which flaring super-function should be used? Select \texttt{1} or \texttt{2} when using function $z_0(R) = |a| + |b|R$, where \texttt{1} select no-fitting and \texttt{2} selects fitting. Selecting \texttt{3} or \texttt{4} sets for tabulated interpolation, with \texttt{3} for no fitting or \texttt{4} for fitting.
    \item[\texttt{flaring\_x}] Either an empty list (e.g. \texttt{[]}) when \texttt{flaring\_function} is 1 or 2. Otherwise a list of positions (expressed in pixels) along $R$, example: \texttt{[0,1,2,3,4,5]}. 
    \item[\texttt{flaring\_y}] Either a two-values list (e.g. \texttt{[$a$,$b$]}) when \texttt{flaring\_function} is 1 or 2. Otherwise a list of flaring $z_0$ strengths, expressed in pixels, along $R$, example: \texttt{[1,1,1.1,1.2,1.4]}.

    \item[\texttt{velocitydispersion\_function}] Which velocity dispersion super-function should be used? Select \texttt{1} or \texttt{2} when using function $\sigma(R) = |a|$, where \texttt{1} select no-fitting and \texttt{2} selects fitting. Selecting \texttt{3} or \texttt{4} sets for tabulated interpolation, with \texttt{3} for no fitting or \texttt{4} for fitting.
    \item[\texttt{velocitydispersion\_x}] Either an empty list (e.g. \texttt{[]}) when \\ \texttt{velocitydispersion\_function} is 1 or 2. Otherwise a list of positions (expressed in pixels) along $R$, example: \texttt{[0,1,2,3,4,5]}. 
    \item[\texttt{velocitydispersion\_y}] Either a one-value list (e.g. \texttt{[$a$]}) when \\ \texttt{velocitydispersion\_function} is 1 or 2. Otherwise a list of velocity dispersion $\sigma$ strengths, expressed in km/s, along $R$, example: \texttt{[10,9,8,7,6]}.

    \item[\texttt{inclination\_function}] Either \texttt{1} for fixed inclination or \texttt{2} for fitting.
    \item[\texttt{inclination\_value}] Value of the inclination $i$, in degrees. Example: \texttt{0.0}.
    \item[\texttt{inclination\_bounds}] Tuple with boundary values for the inclination fitting, in degrees. Example: \texttt{(80.0, 90.0)}.

    \item[\texttt{PA\_function}] Either \texttt{1} for fixed position angle or \texttt{2} for fitting.
    \item[\texttt{PA\_value}] Value of the position angle, in degrees. Example: \texttt{0.0}.
    \item[\texttt{PA\_bounds}] Tuple with boundary values for the position angle fitting, in degrees. Example: \texttt{(-10, 10)}.
    
    \item[\texttt{halolag\_function}] Either \texttt{1} for fixed halo lag or \texttt{2} for fitting.
    \item[\texttt{halolag\_value}] Strength of the halo lag, in km/s per pixel. Example: \texttt{0.2}.
    \item[\texttt{halolag\_bounds}] Tuple with boundary values for the halo lag fitting, km/s per pixel. Example: \texttt{(0,10)}.
    
    \item[\texttt{warp\_function}] Either \texttt{3} for fixed warp or \texttt{4} for fitting.
    \item[\texttt{warp\_theta}] Pitch angle for the warp, in degrees. A pitch angle of $0^\circ$ represents a line-of-sight warp. Example: \texttt{0.0}.
    \item[\texttt{warp\_theta\_bounds}] Tuple with the boundaries on the pitch angle of the warp, for fitting, in degrees. Example: \texttt{(-1.0, 70)}. 
    \item[\texttt{warp\_r}] List of position along $R$ where to interpolate. In pixels. Example: \texttt{[0, 20, 400]}.
    \item[\texttt{warp\_r\_bounds}] List of tuples with the boundaries on the values of \texttt{warp\_r} for fitting, in pixels. Example: \texttt{[(0.0,0.01), (15,40), (400,400.01)]}.
    \item[\texttt{warp\_z}] List of offsets from the central plane, in pixels. Example: \texttt{[0, 0, 0]}.
    \item[\texttt{warp\_z\_bounds}] List of tuples with the boundaries on the values of \texttt{warp\_z} for fitting, in pixels. Example: \texttt{[(0.0,0.0), (0.0,0.0), (0.0,500)]}.
    
    \item[\texttt{dispersiontensor}]The values of the velocity dispersion tensor. Example: \texttt{[1,1,1]}.
    \item[\texttt{dispersiontensor\_free}]Do you want the dispersion tensor to be fitted. Not well supported. Example: \texttt{no}.

    \item[\texttt{beam}]List with beamsizes in arcsec. Should have the same length as \texttt{image}. Example: \texttt{[10]}.
    \item[\texttt{noise}]List with the one-sigma noise levels in each cube given by \texttt{image} in Kelvin. Example: \texttt{[0.79]}.

  \end{description}
\bigskip

  \item[\texttt{[integrator]}] \hfill \\
  \begin{description}
    \item[\texttt{intermediate\_results}]Plot intermediate results during fitting? Example: \texttt{yes}.
    \item[\texttt{accuracy}]Accuracy level of the model. Example: \texttt{1}.
    \item[\texttt{supersampling}]Super-sampling level of the model. Example: \texttt{2}.
  \end{description}
\bigskip

  \item[\texttt{[output]}] \hfill \\
  \begin{description}
    \item[\texttt{radius\_rotationcurve}]Plot the rotationcurve $v_\textrm{rot}(R)$? Example: \texttt{yes}.
    \item[\texttt{radius\_surfacedensity}]Plot the face-on surface density $A_\hi(R)$? Example: \texttt{no}.
    \item[\texttt{radius\_flaring}]Plot the flaring $z_0(R)$? Example: \texttt{yes}.
    \item[\texttt{radius\_velocitydispersion}]Plot the velocity dispersion $\sigma(R)$?\\ Example: \texttt{yes}.
    \item[\texttt{XV\_map}]Plot an integrated PV diagram of the data? Example: \texttt{no}.
    \item[\texttt{radius\_scale}]Controls in which units the radius $R$ is expressed in the previous plots. Options are \texttt{arcsec}, \texttt{kpc} or \texttt{pixels}.
    \item[\texttt{integratedprofile}]Plot an integrated profile of the data? Example: \texttt{yes}.
    \item[\texttt{moment0}]Plot a zeroth moment map? Example: \texttt{no}.
    \item[\texttt{moment1}]Plot a first moment map? Example: \texttt{no}.
    \item[\texttt{moment2}]Plot a second moment map? Example: \texttt{yes}.
    \item[\texttt{moment3}]Plot a third moment map? Example: \texttt{no}.
    \item[\texttt{cube\_visible}]Save the final \hi data cube, expressed in $10^{20}$ atoms/cm$^2$. Example: \texttt{yes}.
    \item[\texttt{cube\_total}]Save the \hi data cube as if it was optically thin, expressed in $10^{20}$ atoms/cm$^2$. Example: \texttt{yes}.
    \item[\texttt{cube\_missing}]Save a cube in which only the hidden atoms are shown, expressed in $10^{20}$ atoms/cm$^2$. Example: \texttt{yes}.
    \item[\texttt{cube\_opacity}]Save the final \hi data cube, expressed in the opacity $\tau$. Example: \texttt{yes}.
    \item[\texttt{cube\_intensity}]Save the final \hi data cube, expressed in Kelvin. Example: \texttt{yes}.
\end{description}
  
\end{description}

\bsp

\label{lastpage}

\end{document}